\documentclass[wcp]{jmlr}

\usepackage{wrapfig}
\usepackage{caption}
\usepackage{enumitem}
\usepackage{amsmath}
\usepackage{amssymb}
\usepackage{bm}
\usepackage{smile}
\newcommand{\bel}{\begin{eqnarray}\label}
\newcommand{\eel}{\end{eqnarray}}
\newcommand{\bes}{\begin{eqnarray*}}
\newcommand{\ees}{\end{eqnarray*}}

\def\benu{\begin{enumerate}}
\def\eenu{\end{enumerate}}

\def\complex{\mathop{{\rm I}\kern-.58em\hbox{\rm C}}\nolimits}

\def\hbar{\overline{h}}

\newcommand{\bfbeta}{{\boldsymbol{\beta}}}
\newcommand{\bfeps}{{\boldsymbol{\epsilon}}}

\newcommand{\bfgamma}{{\boldsymbol{\gamma}}}

\newcommand{\bfzeta}{{\boldsymbol{\zeta}}}

\newcommand{\bfLambda}{{\boldsymbol{\Lambda}}}

\newcommand{\BI}{{\mathbb{I}}}
\newcommand{\BR}{{\mathbb{R}}}

\jmlrvolume{1}
\jmlryear{2021}
\jmlrworkshop{Algorithmic Fairness through the Lens of Causality and Robustness}

\title{Detecting Bias in the Presence of Spatial Autocorrelation}

\author{
\Name{Subhabrata Majumdar}\thanks{Work done while at Data Science and AI Research, AT\&T.}
\addr {Applied ML Research, Splunk}
\Email{smajumdar@splunk.com}
\\
\Name{Cheryl Flynn}
\addr{Data Science and AI Research, AT\&T} 
\Email{cflynn@att.com}
\\
\Name{Ritwik Mitra}\footnotemark[1] 
\Email{rit.stat@gmail.com}
}

\begin{document}
\maketitle

\begin{abstract}
In spite of considerable practical importance, current algorithmic fairness literature lacks technical methods to account for underlying geographic dependency while evaluating or mitigating bias issues for spatial data. We initiate the study of bias in spatial applications in this paper, taking the first step towards formalizing this line of quantitative methods. Bias in spatial data applications often gets confounded by  underlying spatial autocorrelation. We propose hypothesis testing methodology to detect the presence and strength of this effect, then account for it by using a spatial filtering-based approach---in order to enable application of existing bias detection metrics. We evaluate our proposed methodology through numerical experiments on real and synthetic datasets, demonstrating that in the presence of several types of confounding effects due to the underlying spatial structure our testing methods perform well in maintaining low type-II errors and nominal type-I errors.
\end{abstract}


\section{Introduction}
\label{sec:intro}
In recent literature, fairness concerns have been studied in machine learning (ML)-based criminal justice systems, credit scoring, and facial recognition among others.  In comparison, ML fairness in problems involving \textit{spatial data} has been studied to a relatively lesser extent.  Idiosyncrasies associated with spatial data such as spatial autocorrelation can introduce an underlying dependence structure in the data. As a result, existing fairness definitions and techniques designed for independent data may no longer apply.  In this paper, we present a first attempt towards addressing this gap by formalizing a quantitative approach for detecting bias in spatial data applications. 


Deployment locations of services such as bike-share locations, cell towers, and user-stops for location-based gaming may need to be constrained by minimum and maximum distances between adjacent pairs or groups of locations. Potential reasons for imposing such constraints may range from an overall goal for optimal coverage and accessibility, to limitations of the underlying technology (e.g. 5G cellular networks). Such service locations, especially when deployed through for-profit corporations, may be selected using ML in a manner that attempts to maximize engagement with or subscription to the service while satisfying any necessary spatial constraints. To assess the bias in such applications, one might consider the association between selected locations and their corresponding demographics. However, for historic and societal reasons, distribution of different demographic groups is also usually spatially autocorrelated~\citep{segre}.  These underlying spatial patterns in the data can result in a confounding effect that can invalidate standard bias detection procedures.

\begin{figure}[t]
\centering

\subfigure[Confounding effects of underlying spatial structure]
{\label{fig:causal}
\includegraphics[width=.65\linewidth]{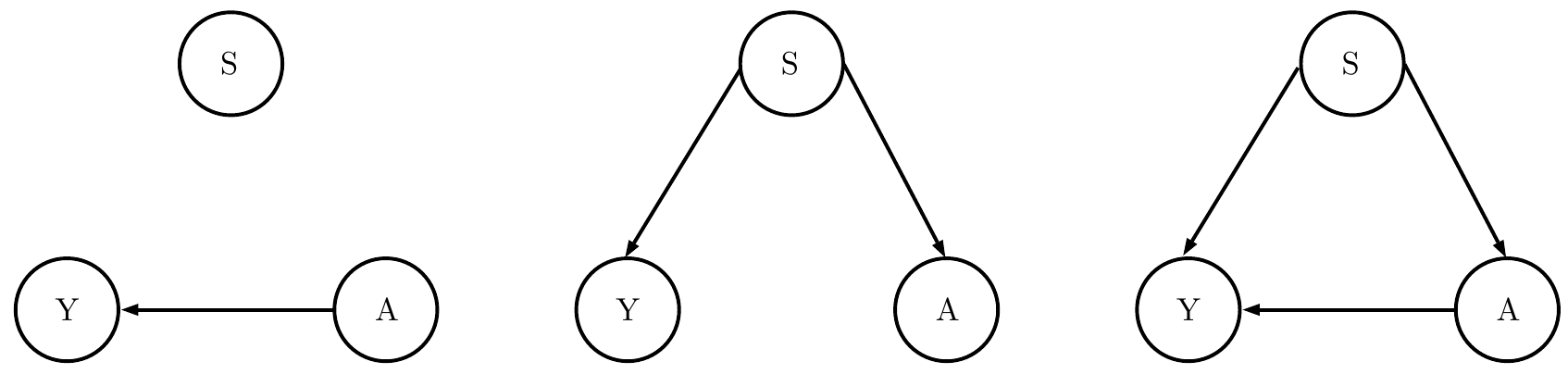}
}
\subfigure[Simple Tile Example]
{\label{fig:exampleTiles}
\includegraphics[width=.28\linewidth]{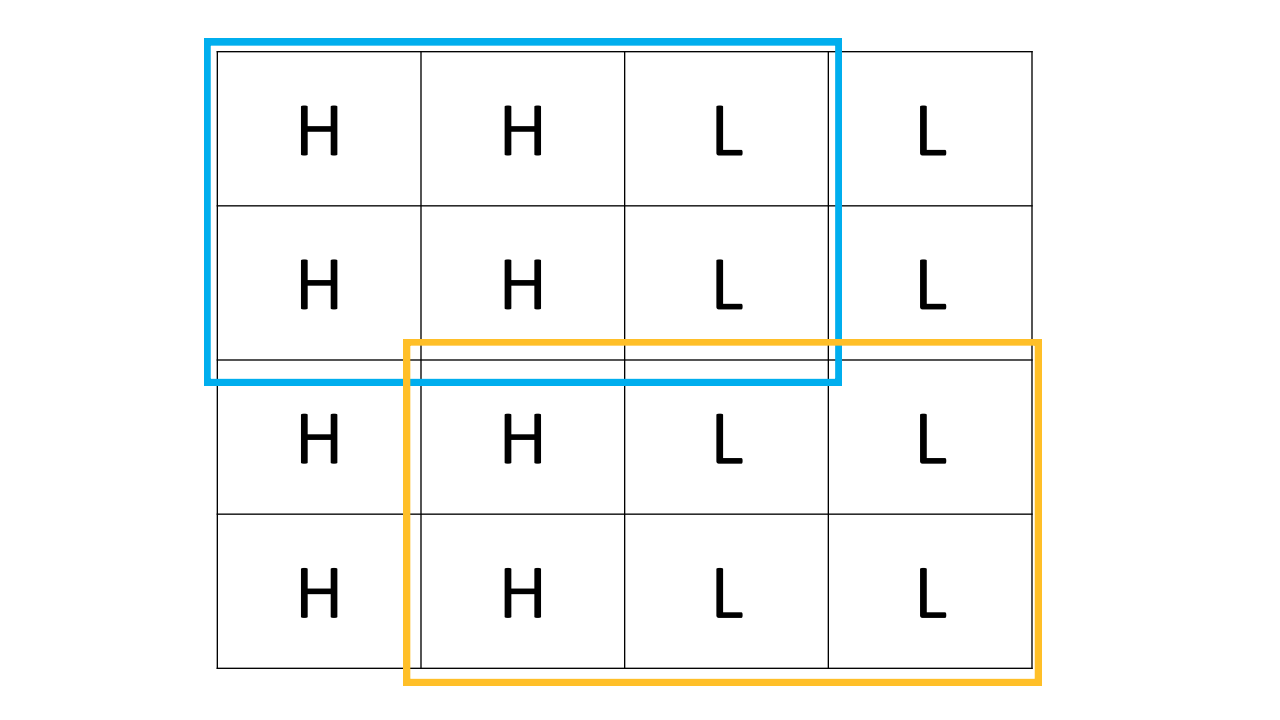}
}

\caption{In presence of spatial autocorrelation, associations between the outcome feature $Y$ and sensitive feature $A$ get confounded with potential associations with $S$ (panel a). A simple tile example demonstrates that this effect translates to increased likelihood of detecting bias in presence of spatial autocorrelation (panel b).}
\label{fig:exampleFig}
\end{figure}

Consider the simple tile setup in Figure~\ref{fig:exampleTiles}.  Each tile is assigned a value of High ($H$) or Low ($L$) to represent a demographic attribute such as income or percent minority.  Suppose we randomly select 6 tiles and compute the disparate impact ($DI$) bias metric, where
\[
f = \frac{|Selected \cap H|/|H|}{|Selected \cap L|/|L|},
\quad DI = \max(f, 1/f).
\]
Assuming $DI > 1.2$ indicates a biased selection process, let us compare the likelihood of detecting bias in two scenarios.  In scenario 1, we impose no spatial constraints.  Then the number of $H$ and $L$ tiles follows a hypergeometric distribution and the probability distribution of $DI$ can be computed exactly, where bias is detected 60.84\% of the time.  In scenario 2, we only select $2\times3$ rectangles. There choices can have either four or two H tiles (blue or yellow border in Fig.~\ref{fig:exampleTiles}, respectively), so that $DI = 2$ always, and bias is detected with 100\% probability.  Thus, the combination of the presence of spatial autocorrelation in the demographic attribute and the selected tiles alters the distribution of the bias metric.  
%


\begin{wrapfigure}[14]{r}{0.5\textwidth}
\centering
\includegraphics[width=.24\textwidth]{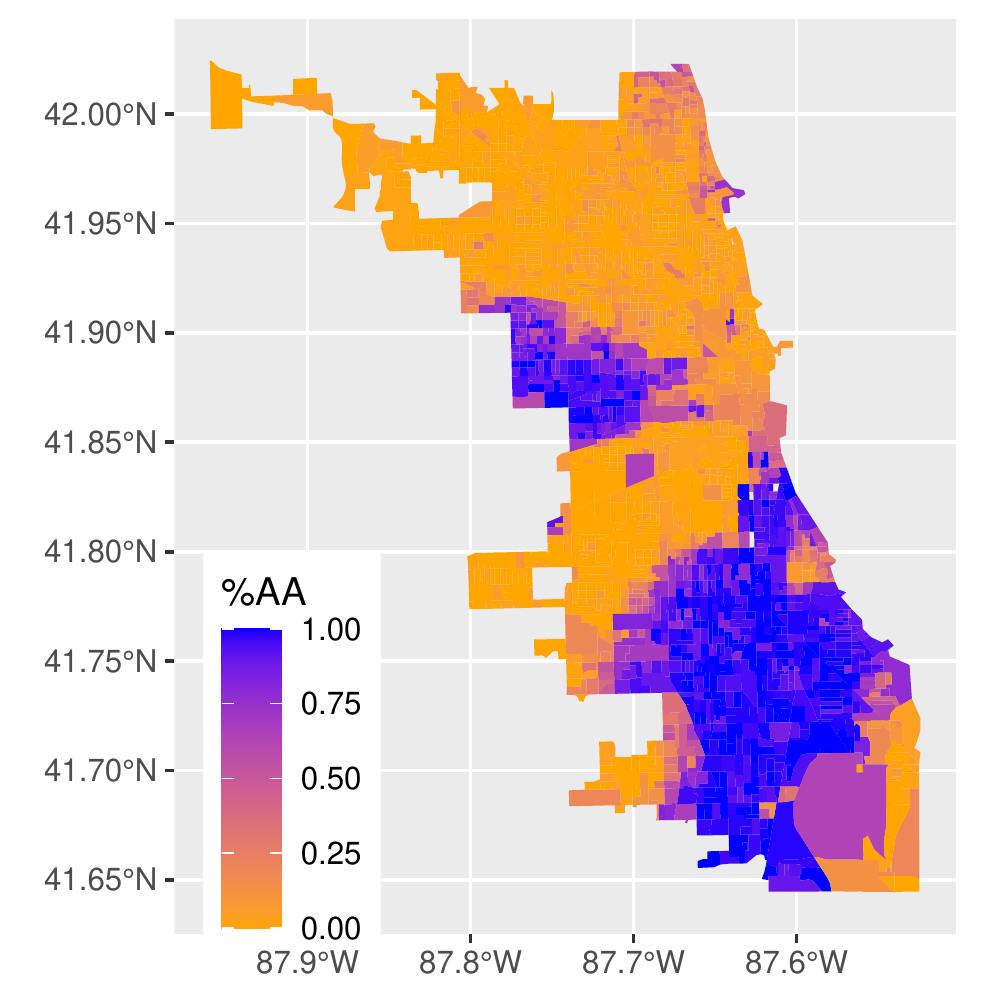}
\includegraphics[width=.24\textwidth]{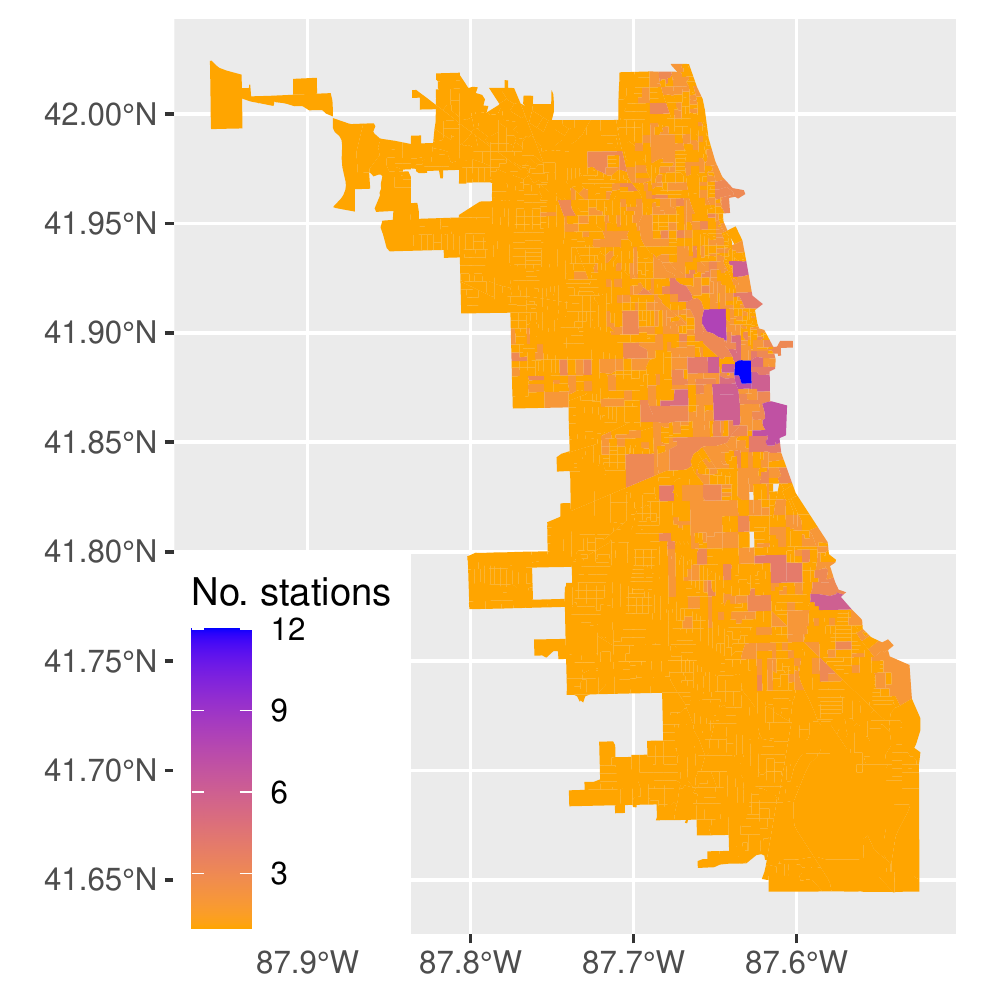}
\caption{Spatial distribution of bikeshare stations in Chicago (left) is largely complementary to that of the percentage of African Americans (right) at census tract level.}
\label{fig:chicmap}
\end{wrapfigure}

Formalizing the above notion, in the presence of a spatial dependency pattern $S$ between partitions of a geographic area, the estimation of any association or cause-effect relationship between a target variable, $Y$, and a sensitive attribute, $A$, may be confounded by $S$. As demonstrated in the causal diagram in Fig.~\ref{fig:causal}, there are multiple subcases of the above scenario, which may result in overestimation of the degree of association between $Y$ and $A$ through any spurious correlation either or both of them might have with $S$.  From left to right, these diagrams capture three cases: (1) direct relationship between $Y$ and $A$, (2) indirect relationship between $Y$ and $A$ through $S$, and (3) direct relationship between $Y$ and $A$ with additional association through $S$.  Bias detection methods should account for these relationships to avoid a potentially erroneous conclusion.

\paragraph{Our contributions}
We provide a framework to evaluate bias concerns in spatial data, specifically in scenarios where confounding effects of spatial autocorrelation influence the outcomes of interest, sensitive attribute values, or both. Under this setup, we aim to evaluate a number of scenarios regarding the effect of an underlying spatial structure: (a) the detection of a common spatial effect, (b) determining whether these effects are equivalent in magnitude, and (c) testing for association between $Y$ and $A$ while adjusting for this spatial effect. We propose methods that provide answers to them (Sec.~\ref{sec:setup}), and evaluate these methods in a number of data settings, including the Chicago bikeshare data in Fig.~\ref{fig:chicmap}.

\paragraph{Related work}
Spatial fairness is an important problem with a myriad of applications that affect the populace in direct and tangible ways. Spatial distribution of resources impact the accessibility (or lack there of) of crucial private and public goods, such as housing, transportation, schools, and healthcare. Historically, systemic, intentional and even unintentional biases have affected distribution of these resources. This encompasses redlining for lending applications~\citep{corbett2018measure}, and inequity concerns in bikeshare problems \citep{goodman2014,natco2015,ogilvie2012}, healthcare solutions \citep{guagliardo2004spatial,jin2015spatial}, COVID-19 vaccine enrollment \citep{covid1,covid2}, and Pok{\'e}mon GO \citep{colley2017geography}.

In spite of the very real needs, to the best of our knowledge there are no known methods to detect and mitigate spatial fairness issues. 
Till now, most algorithmic fairness methods \citep{MehrabiEtal19} deal with obtaining {\it point estimates} of any association between the outcome $Y$ and sensitive attribute $A$, or mitigating this association while estimating $Y$, and not the {\it significance} of such association. While some very recent papers do propose hypothesis tests for evaluating fairness~\citep{BlackEtal20,dicchio20}, they assume the data samples analyzed to be independent. As we have seen in the motivating example, the presence of an underlying dependency pattern can result in such tests conflating this dependence with the presence of actual demographic bias.



\section{Methodology}
\label{sec:setup}
Consider $n$ location units $\{l_{i}\}_{i=1}^{n}$, the spatial dependency between which is quantified by a weight matrix $\Wb$. For location unit $i$, we have non-sensitive input features $\bX_{i} \in \BR^{p}$--giving rise to data matrix $\Xb = (\bX_1, \ldots, \bX_n)^T \in \BR^{n\times p_{1}}$, a (discrete or continuous) output feature $\by\in\BR^{n}$ indicating the allocation or presence of a resource at location $i$, and $\Ab = (\bA_1, \ldots, \bA_n)^T \in \BR^{n\times p_2}$ as the data on sensitive input features.  
The realized value of a random vector is denoted by lowercase; e.g. $\bX_{i}$, takes the value $\bx_{i}$. For simplicity, we assume a single sensitive  attribute $A$, i.e. $p_2=1$, hereafter.

Under the above setup, we aim to answer three questions: (Q1) Are both $Y$ and $A$ significantly associated with the common weight matrix $\Wb$? (Q2) Given that they are, are their strength of associations with $\Wb$ similar? (Q3) Can we adjust for spatial association to measure the true degree of association between $Y$ and $A$?
\subsection{Testing for common autocorrelation}
Moran's I \citep{moran1950notes} is a well-known statistic to test for global dependence in spatial data. Given realizations $\bz = (z_1, \ldots, z_n)^T$ of a continuous random variable $Z$ and a weight matrix $\Wb = ((w_{ij}))$, Moran's I is defined as
$$ I = \frac{n}{\bar w} \frac{\sum_{i=1}^n \sum_{j=1}^n w_{ij} (y_i - \bar y)(y_j - \bar y)}
{\sum_{i=1}^n (y_i - \bar y)^2};\quad
\bar w = \frac{1}{2} \sum_{i,j} (w_{ij} + w_{ji}), \bar y = \frac{1}{n} \sum_i y_i. $$
Under the null hypothesis (Appendix~\ref{app:hyp}) that no spatial autocorrelation exists, a centered and scaled version of $I$ is asymptotically standard Gaussian \citep{ShapiroHubert79,OneilRedner93}. Recently, \cite{LeeOgburn17} generalized the Moran's I for realizations of a categorical random variable, and proposed an analogous test statistic. When observations are binary, standardized versions of Moran's I and this statistic above are equivalent, and asymptotically standard Gaussian \citep{LeeOgburn17}. Asymptotic Gaussian tail-bounds or permutation tests can be used to obtain $p$-values of either test statistic.

Assume that the standardized Moran's I calculated using the size-$n$ sample $\by$ (or $\bA$) and $\Wb$ follow the (non-asymptotic) location family distribution
$F(\cdot,\mu_y)$ (or $F(\cdot,\mu_a)$) with mean parameter $\mu_y$ (or $\mu_a$). Then, to answer the question (Q1) above, we test the following:
\begin{align*}
H_{0y}^1: \mu_y = 0 \text{ vs. } H_{ay}^1: \mu_y \neq 0
\quad\text{and}\quad
H_{0j}^1: \mu_a = 0 \text{ vs. } H_{aa}^1: \mu_a \neq 0.
\end{align*}
Specifically, we are interested in knowing whether the scenario when both the alternate hypotheses hold, i.e. both $y$ and $A$ exhibit significant autocorrelation. Thus we test for the combined null $H_0^1 = H_{0y}^1 \bigcup H_{0a}^1$ against the combined alternative $H_a^1 = H_{ay}^1 \bigcap H_{aa}^1$. To answer question (Q2), we test $H_0^2 = H_0^{21} \bigcup H_0^{22}$ against $H_a^2 = H_a^{21} \bigcap H_a^{22}$, where
\begin{align*}
H_{0}^{21}: \mu_y - \mu_a \geq \delta \text{ vs. }
H_{a}^{21}: \mu_y - \mu_a < \delta \quad\text{and}\quad
H_{0}^{22}: \mu_y - \mu_a \leq -\delta \text{ vs. }
H_{a}^{22}: \mu_y - \mu_a > -\delta,\\
\end{align*}
for some fixed $\delta>0$, and with the additional assumption that $\mu_y \neq 0, \mu_a \neq 0$. Guidance for useful values of $\delta$ may be determined by target thresholds of bias metrics (Appendix~\ref{app:metrics}), such as the disparate impact thresholds of $(0.8,1.2)$ based on the EEOC four-fifths rule \citep{eeoc}. We take $\delta = 1$ in our experiments to account for small differences that may occur under null between the two standardized Moran's I statistics due to sampling noise.

Working under the above setup--known as the Intersection-Union (IU) principle~\citep{Berger82,BergerHsu96}--we combine test statistics for each of the pairs of sub-hypotheses above using the concept of {\it Pivotal Parametric Products}:
\begin{definition}[\cite{Sengupta07}]
For the problem of testing for a (possibly vector-valued) parameter $\theta$ using a union null $H_0 \equiv \bigcup_{s=1}^S H_{0s}$, a Pivotal Parametric Product (P3) is defined as any function $\eta \equiv g(\theta)$ such that $H_0$ holds if $\eta = 0$.
\end{definition}
Testing for $H_0: \eta=0$ can be done using an (unbiased or consistent) estimator of $\eta$. In our context, we define the P3 functions to test for $H_0^1$ vs. $H_a^1$ and $H_0^2$ vs. $H_a^2$ as $P_1 = \min(|\mu_y|, |\mu_a|); \quad P_2 = \left| (\mu_y-\mu_a)^2 - \delta^2 \right|$. Assuming that the individual Moran's I test statistics are called $I_y$ and $I_a$, respectively for $y$ and $A$, the corresponding test statistics will be: $T_1 = \min(|I_y|, |I_a|); \quad T_2 =  \left| (I_y-I_a)^2 - \delta^2 \right|$. Denoting the lower-$p$ ($p \in [0,1]$) tail of the probability distribution of a statistic $T$ by $T_p$, rejection regions at level $\alpha$ for the above tests are characterized by large values of $T_1$ and small values of $T_2$, specifically by the sets $\{T_1 \geq T_{1,1-\alpha} \}$ and $\{T_2 \leq T_{2,\alpha} \}$, respectively. 


We use permutation test (Appendix~\ref{app:permtest}) to obtain $p$-values for above tests. We construct the null distributions for each test statistic by calculating them over randomly permuted samples, then compute the tail probabilities of the statistics above with respect to the respective distributions.

\subsection{ESF adjustment}
\label{subsec:adj-proc}
Eigenvector Spatial Filtering (ESF) \citep{dray2006,griffith2004,GRIFFITH2019xvii} is a popular method to remove underlying spatial dependencies from spatial observations. It fits a linear regression on that feature, with eigenvectors of $\Wb$ and a number of covariates as input features. Residuals from this regression can be used as dependency-free analogues of the original observations for further analysis. Formally, this {\it spatial autoregression} model assumes $
    \by = (\Ib_n - \rho \Wb)^{-1} (\Xb \bfbeta + \bfeps); \quad
    \bfeps \sim N(0, \sigma^2 \Ib_n), \sigma>0 $,
where $\rho \in [-1,1]$ denotes spatial autocorrelation, and $\bfbeta \sim \BR^p$ quantifies the (linear) effect of $\bX$ on $\bY$. Given that the spatial dependency in $\bY$ is entirely due to the top $k$ few eigenvectors of $\bW$, Consider now the spectral decomposition $\Wb = \Eb \bfLambda \Eb^T$, where $\Eb \in \BR^{n \times n}$ is orthogonal and $\bfLambda \in \BR^{n \times n}$ is diagonal. Using straightforward algebra \citep{GRIFFITH2019xvii}, this model simplifies to
\begin{align}
    \by = \rho \Eb \bfLambda \Eb^T \by + \Xb^T \bfbeta + \bfeps \simeq \rho \Eb_k \bfgamma_k + \Xb^T \bfbeta + \bfzeta, \label{eqn:eigk}
\end{align}
where $\bfgamma_k$ estimates the effect of the top $k$ eigenvectors, and $\bfzeta \sim N(0, \sigma_1^2 \Ib_n), \sigma_1>0$. The least square estimates of $\bfgamma_k, \bfbeta$ can be plugged in to obtain a `sanitized', decorrelated version of $\by$:
$ \tilde \by = \by - \rho \Eb_k \hat \bfgamma_k + \Xb \hat \bfbeta. $

To address (Q3) above, we use ESF to learn features that capture the spatial structure in $Y$ and $A$, then use these features to adjust the bias detection procedure. Depending on whether the feature of interest is continuous or discrete, we use a squared error or logistic loss--with $\ell_1$ penalization to automatically choose the best subset of eigenvectors (Appendix~\ref{app:esf}). To choose the optimal tuning parameter $\lambda$, we use Akaike Information Criterion (AIC) and Bayesian Information Criterion (BIC).
\section{Experiments}
\label{sec:simulations}

\subsection{Synthetic data: testing for common autocorrelation}
We take the centroid locations of $n=2080$ census block groups in the city of Chicago, construct a weight matrix $\Wb$ using the inverse of distances between tracts $i,j$ as the weight $w_{ij}, 1 \leq i, j \leq n$, and generate samples of the random variables $(Y, \bX, A)$. When required, we use 1000 permutations to generate approximate null distributions, and compute all empirical rejection rates using $1000$ Monte Carlo runs. We evaluate the performance of the IU-combination tests in three different scenarios.

\paragraph{Association by autocorrelation}
We first focus on the case where both the response $Y$ and the sensitive attribute $A$ are independently correlated with $\Wb$. We start with standard Gaussian errors $\bfeps_1, \bfeps_2 \sim N(0,\Ib_n)$, and given a value of spatial autocorrelation $\rho$, generate spatially lagged features:
$ \by = ( \Ib - \rho \Wb)^{-1} \bfeps_1,
\ba = ( \Ib - \rho \Wb)^{-1} \bfeps_2. $

\begin{wrapfigure}[12]{r}{.5\textwidth}
    \centering
    \includegraphics[width=.5\textwidth]{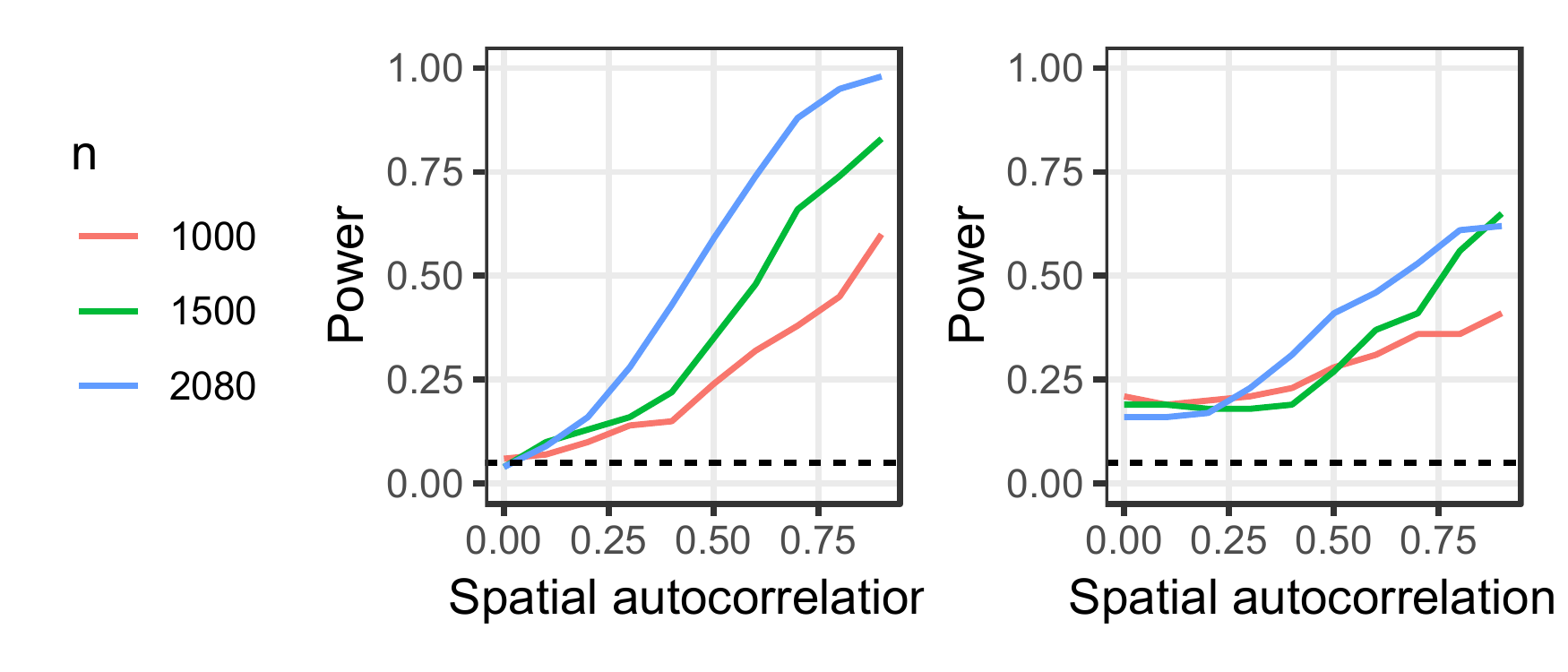}
    \caption{Powers for $T_1$ (left) and $T_2$ (right) under association by correlation. Dotted lines indicate size $\alpha = 0.05$.}
    \label{fig:fig1}
\end{wrapfigure}

We repeat the experiment for $\rho = 0, 0.1, \ldots, 0.9$, and different sample sizes, by either randomly choosing 1000 and 1500 census tracts or using all 2080 census tracts, to quantify how the degree of spatial autocorrelation and sample size affect the power (i.e. Type-II error) of the testing procedure.


We summarize the results in Fig.~\ref{fig:fig1}. As $\rho$ increases, the test $T_1$ is able to detect the existence of a non-zero $\rho$ with higher power, and also maintains nominal size, as the value of $T_1$ is exactly 0.05 for $\rho = 0$. For $T_2$, note that rejection implies inferring the equality of $\rho$ across $\by$ and $\ba$. Thus Fig.~\ref{fig:fig1} indicates that our permutation test is also able to infer this with higher power as $\rho$ grows larger. Note that for $T_2$ the alternate hypothesis implies inferring the equality of spatial autocorrelation. This is the case here, so the test rightly maintains rejection rates above $0.05$ for all values of $\rho$.

\paragraph{Association and autocorrelation}
In this situation $Y$ and $A$ are associated independently, and one or both are spatially autocorrelated. To represent such scenario, we generate synthetic data using the two settings, starting with standard Gaussian errors $\bfeps_1, \bfeps_2$ as before:
\begin{eqnarray}
\ba_0 = \bfeps_2, \by = 5 \ba_0 + \bfeps_1, \ba = ( \Ib - \rho \Wb)^{-1} \ba_0,\label{eqn:s21}\\
%
%
\text{and}\quad \ba = ( \Ib - \rho \Wb)^{-1} \bfeps_2, \by = 5 \ba + \bfeps_1.
\label{eqn:s22}
\end{eqnarray}
In the first case only the sensitive attribute is spatially autocorrelated, while in the second situation both $Y$ and $A$ are.

\begin{figure}[t]
\centering
\includegraphics[width=.49\columnwidth]{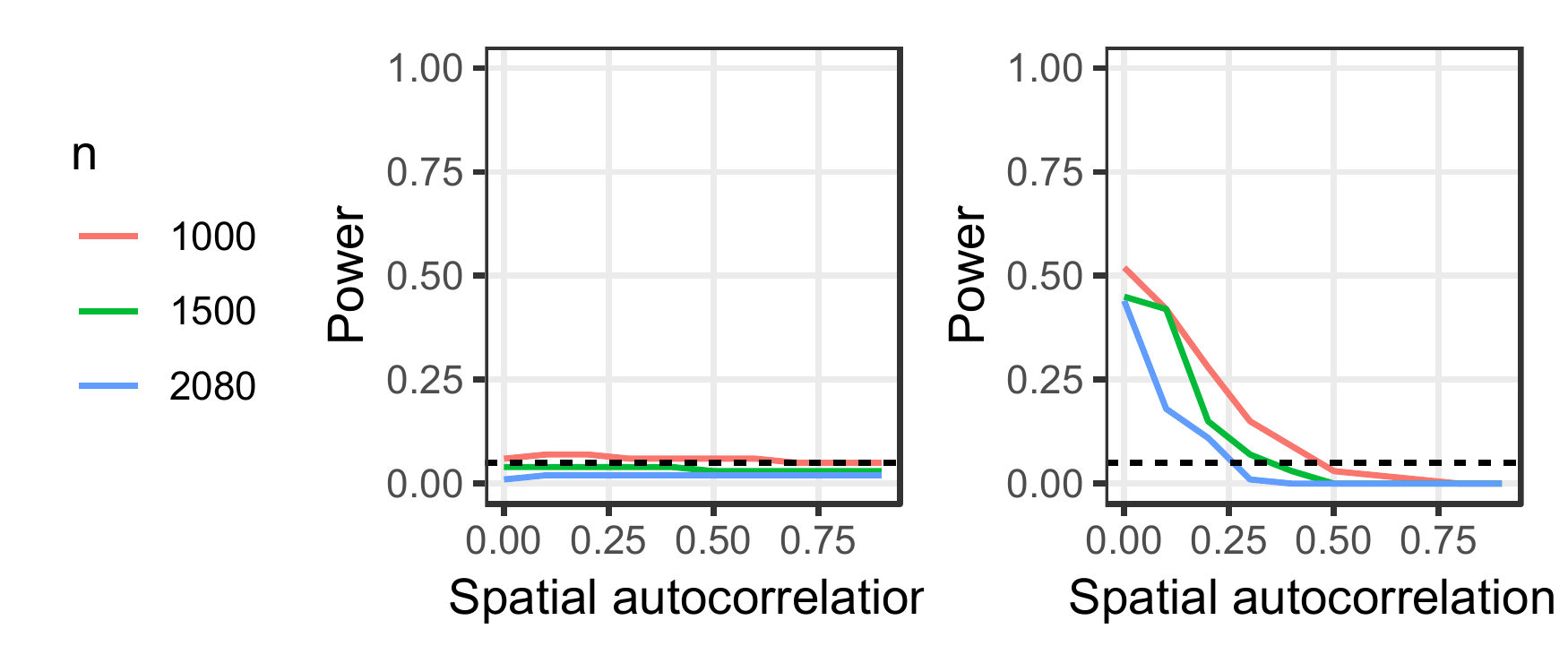}
\includegraphics[width=.49\columnwidth]{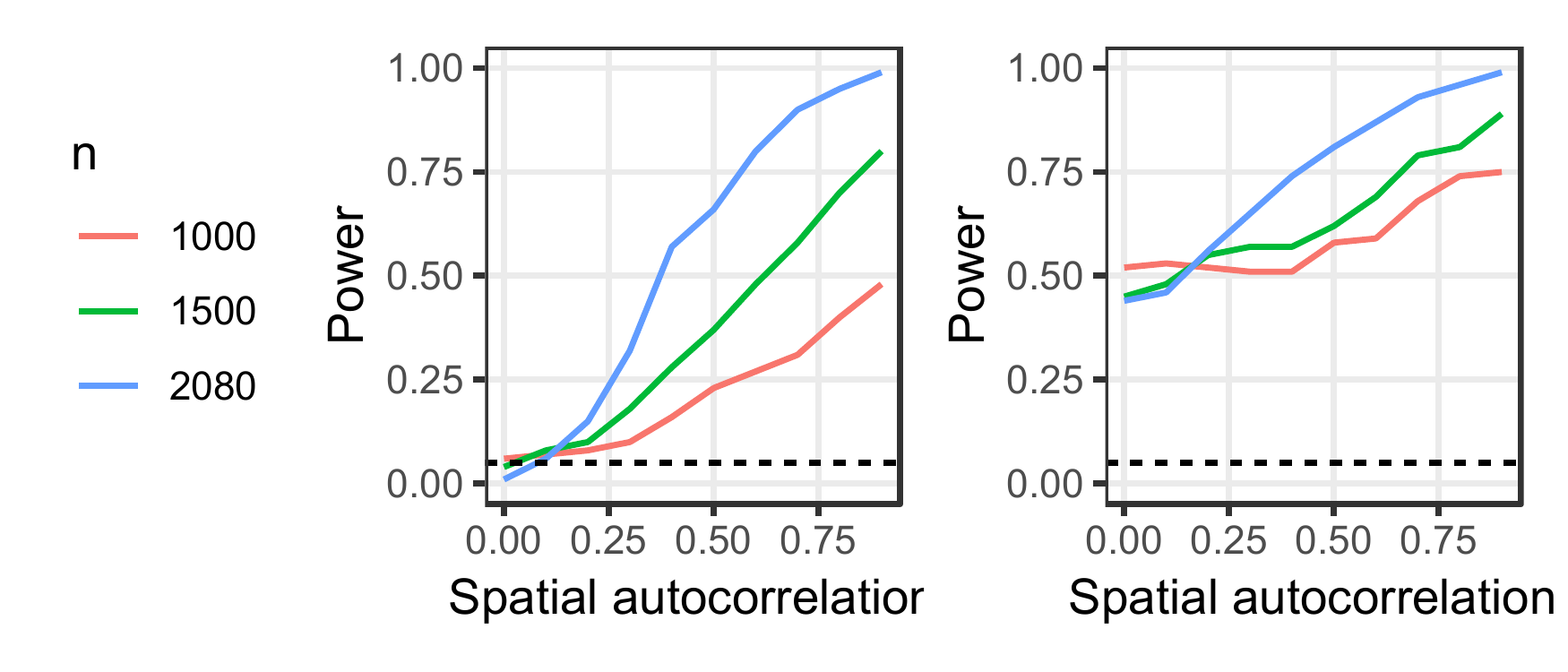}
\caption{Powers for $T_1$ (left) and $T_2$ (right) under association and correlation. Top and bottom rows correspond to settings \eqref{eqn:s21} and \eqref{eqn:s22}, respectively.}
\label{fig:fig2}
\end{figure}

We summarize the results in Fig.~\ref{fig:fig2}. The statistic $T_1$ geared towards finding evidence of a common level of spatial autocorrelation (rightly) maintains nominal size across different values of $\rho$ in the first case above, and increasingly becomes more powerful for higher $\rho$ in the second case. The statistic $T_2$ maintains nominal rejection rates for moderate to high $\rho$ in the first case, giving evidence that there is a mismatch of autocorrelation magnitudes for $Y$ and $A$. In the second case, it however maintains higher rejection rates.

\paragraph{Proxy autocorrelation}
We now consider the third scenario, of $Y$ being associated with {\it non-sensitive} attributes, i.e. components of $\bX$, but not $A$. However $X$ and $A$ are spatially autocorrelated with the same $\Wb$:
$ \bx = ( \Ib - \rho \Wb)^{-1} \bfeps_2, \by = 5\bx + \bfeps_1;
\ba = ( \Ib - \rho \Wb)^{-1} \bfeps_2$.
The results for this setting (not shown) are very similar to Fig.~\ref{fig:fig1}. This is because setting 3 can be seen as the noisy version of setting 1, where we were essentially testing for a common source of autocorrelation between $\bx$ and $\ba$, as compared to $\by$ and $\ba$ here.

\subsection{Synthetic data: adjusting for autocorrelation}
We consider two different values of the underlying spatial autocorrelation, and evaluate the ESF-based testing methods for discrete and continuous attributes across above settings.

\paragraph{Association by autocorrelation}
For the continuous setup, we consider the model
$ \by = ( \Ib - \rho_1 \Wb)^{-1} \bfeps_1,
\ba = ( \Ib - \rho_2 \Wb)^{-1} \bfeps_2$,
and compare error rates of asymptotic tests using the correlation coefficient before and after ESF-adjustment. For the discrete setup, we take $\by \leftarrow \BI(\by > 0), \ba \leftarrow \BI(\ba > 0)$, and compare using the disparate impact metric. In this case we obtain the null distribution using a permutation procedure. In both cases, we consider the range of values $\rho_1 = 0.9, \rho_2 = -0.9, -0.85, \ldots, 0.85, 0.9$.
\begin{figure*}[t]
\centering
\subfigure[Association by autocorrelation]
{\label{subfig:esf1}
\includegraphics[width=.23\textwidth]{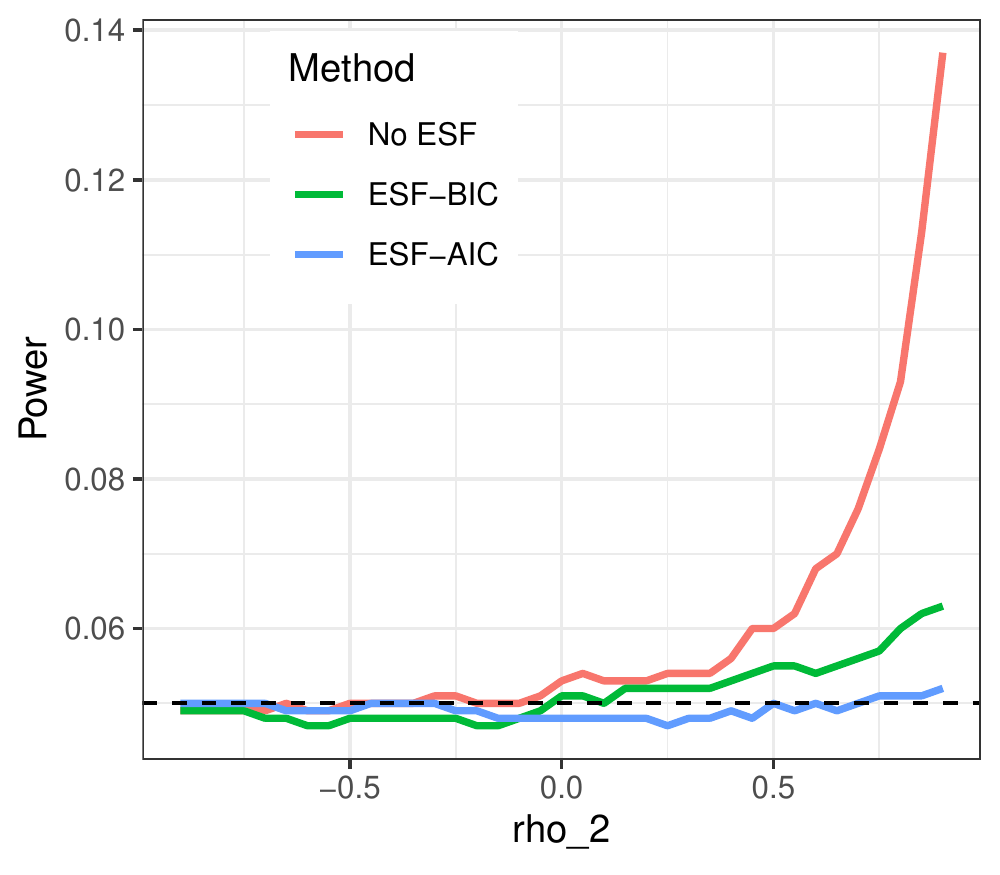}
\includegraphics[width=.23\textwidth]{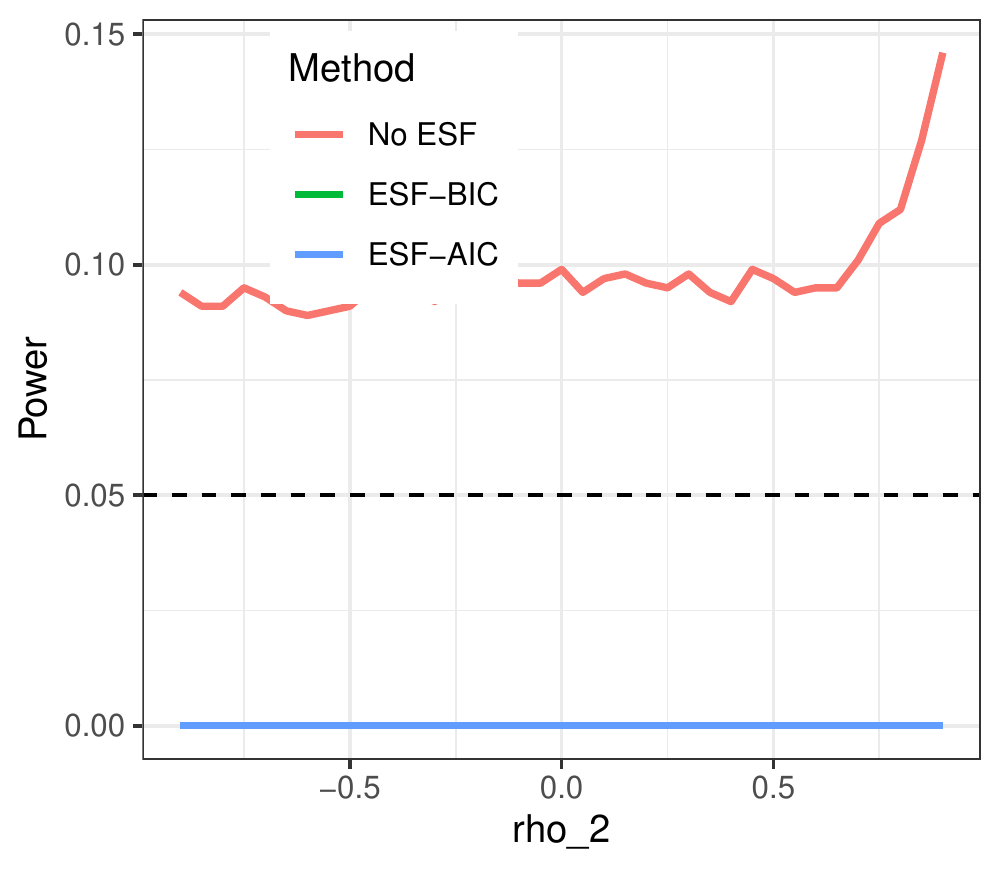}
}
\subfigure[Association and autocorrelation]
{\label{subfig:esf2}
\includegraphics[width=.23\textwidth]{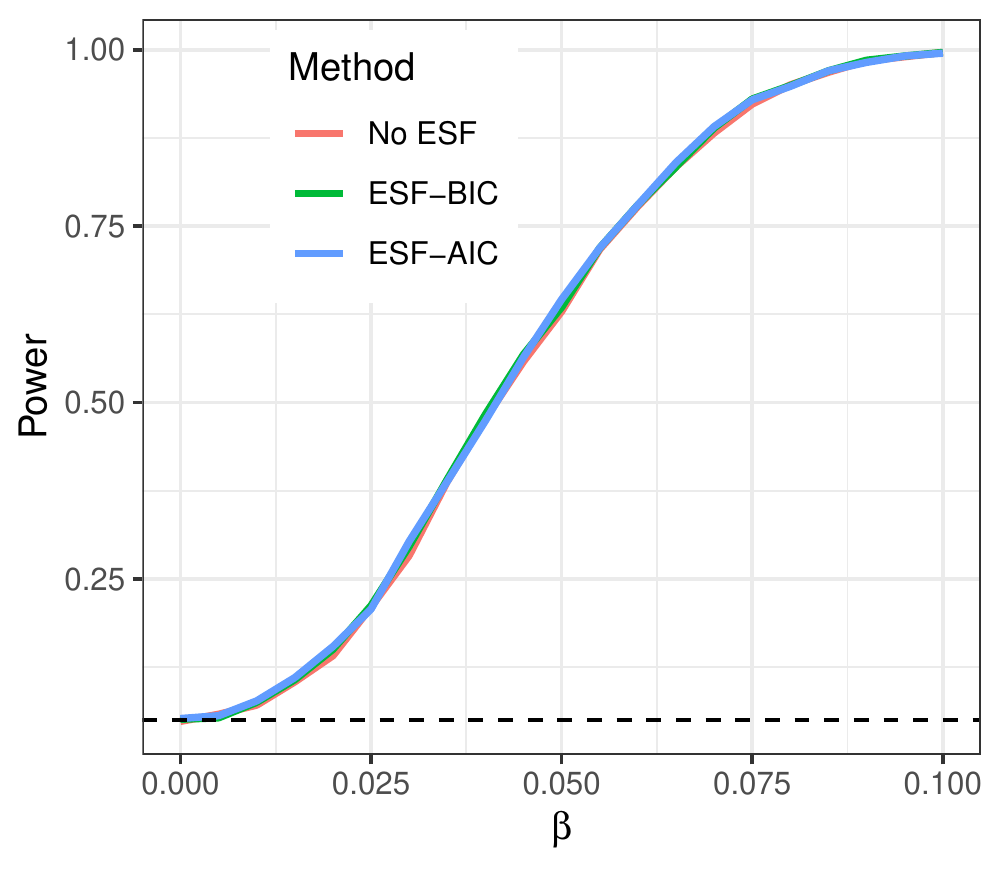}
\includegraphics[width=.23\textwidth]{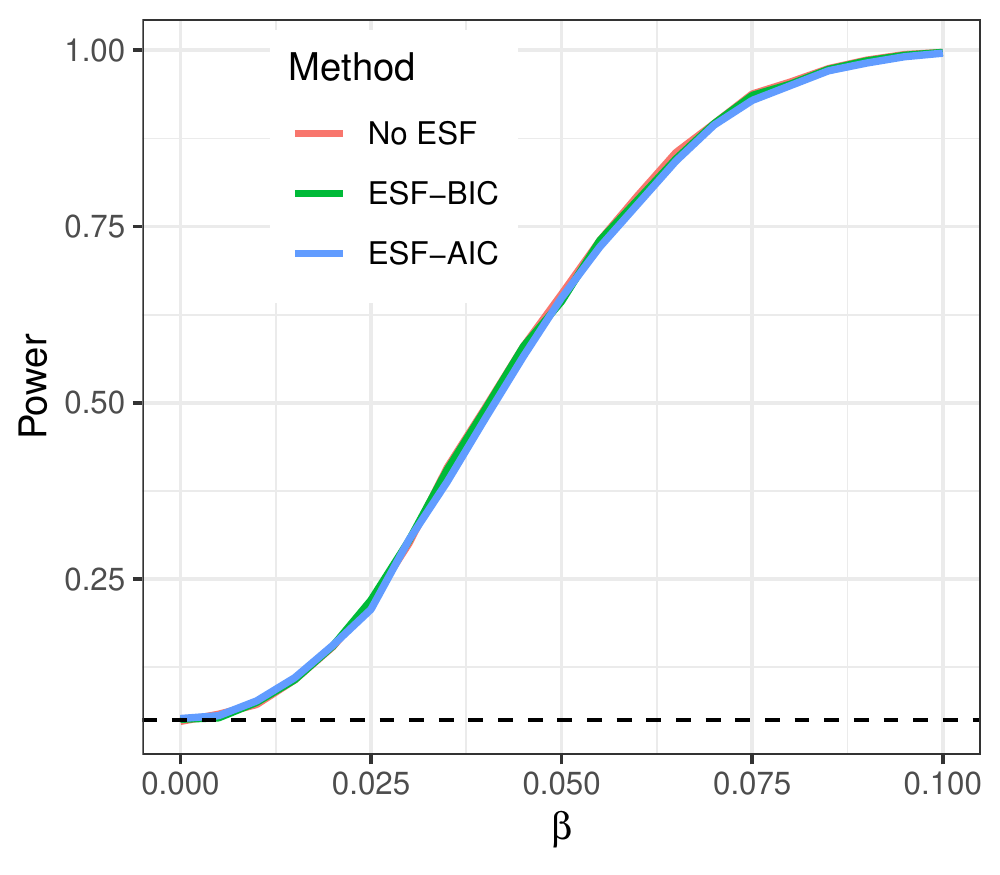}
}
\caption{In each subfigure, rejection rates are given for the ESF-adjusted tests for (left panels) continuous or (right panels) discrete $Y,A$.}
\end{figure*}
We summarize the results in Fig.~\ref{subfig:esf1}. Without ESF correction, rejection rates stay above the nominal 0.05 level across all values of $\rho_2$, but come down below nominal level after correction. The results due to proxy autocorrelation are largely similar to this setting, so we omit those results.

\paragraph{Association and autocorrelation}
In this setup, we consider the case when both $Y,A$ are discrete. We consider the two subcases:
%
\begin{eqnarray*}
\ba_0 = \bfeps_2, \ba = \BI(( \Ib - \rho \Wb)^{-1} \ba_0 > 0),
\by = \BI(\beta \ba_0 + \bfeps_1 > 0),\\
%
%
\text{and}\quad
\ba = \BI(( \Ib - \rho \Wb)^{-1} \bfeps_2 > 0), \by = \BI( \beta \ba + \bfeps_1 > 0).
\end{eqnarray*}
We present the results in Fig.~\ref{subfig:esf2}, calculated on a range of values for $\beta$, and fixing $\rho=0.9$---with outputs for the first case on the left and second case on the right. It is evident from the results that the rejection rates remain the same pre- and post-filtering as expected.

\subsection{Chicago Bikeshare data}
\label{bikeshare}
\begin{table*}[t]
\centering
\scalebox{.8}{
    \begin{tabular}{l|lllll}
    \hline
    $\Wb$              & $\BI$(no. stations $\geq$ 1)         & \%AA       & Med. inc. & $\BI$(\%AA $ \geq 30$) & $\BI$(Med. inc. $\geq $50k) \\ \hline
    Exp, $V = 1$     & 96.06 (0)      & 594.88 (0)   &  211.94 (0)   &  529.95 (0)  &  197.46 (0)                     \\
    Exp, $V = 0.1$   & 107.2 (0)      &  516.73 (0)  &  234.19 (0)   &  460.11  (0) & 195.15 (0)                      \\
    Exp, $V = 0.01$  & 61.01 (0)      &  213.45  (0) & 170.67 (0)    &  192.4  (0)  & 123.44 (0)                      \\
    Exp, $V = 0.001$ & 32.41  (0.001) & 34.93  (0)   &  34.95 (0)    &   34.69 (0)  & 34.11 (0)                       \\
    1-hop neighbor & 2.28 (0.011)   &  10.22 (0)   &    5.43 (0)   &    9.2 (0)   & 4.85 (0)                        \\
    2-hop neighbor & 1.87 (0.032)   &   12.67 (0)  &    8.59   (0) & 11.66 (0)    & 6.11 (0)                        \\\hline
    \end{tabular}}
    \caption{Standardized Moran's I statistics for the Chicago bikeshare data, with permutation test $p$-values in parentheses.}
    \label{table:hyptable}
\end{table*}
The Chicago bikeshare data \citep{bikedata} contains latest information on the location and other neighborhood-related attributes of all bikeshare stations in the city of Chicago. We join this data with the 2018 ACS Census data to obtain two sensitive features at census tract level---percentage of African American population (\%AA), and median income. As we had seen in Fig.~\ref{fig:chicmap}, there is evidence of potential spatial fairness issues in this context. We perform Moran's I-based tests for a number of features to determine their dependence on the weight matrix $\Wb$. We consider several choice of $\Wb$: spatial cross-covariance matrices with exponential covariance function having scaling parameter $V = 10^i; i = 0, -1, -2, -3$, and adjacency matrices indicating 1-hop and 2-hop neighbors of block groups. The results in Table~\ref{table:hyptable} show that both sensitive features and the output feature are spatially auto-correlated. Based on the above single-feature Moran's I, the IU combination test statistics $T_1$ calculated turns out to be significant for all sensitive features, but $T_2$ does not. This demonstrates the presence of a common spatial factor, but with uneven magnitudes of effects on the bikeshare indicator versus any of the sensitive features.

\begin{wraptable}[9]{r}{0.6\textwidth}
\centering
\scalebox{.8}{
    \begin{tabular}{llll}
    \hline
    Feature pair                  & No ESF        & ESF-BIC        & ESF-AIC        \\\hline
    \%AA vs. No. of stations      & \textbf{0.007}         & 0.27           & 0.35           \\
    Med. Inc. vs. No. of stations &\textbf{4.4e-16}  & 0.999          & 0.99          \\
    \%AA vs. Med. Inc.            & \textbf{5.8e-129} &   \textbf{2.6e-5} &   \textbf{1.3e-8} \\\hline
    \end{tabular}}
    \caption{Pairwise $p$-values from asymptotic correlation tests with and without ESF. Bold indicates significant evidence of rejecting null (at level 0.05).}
    \label{tab:chictable}
\end{wraptable}
We perform ESF on three census tract-level features: number of bikeshare stations, \%AA, and median income, based on eigenvectors of the inverse-distance weighted $\Wb$. Table~\ref{tab:chictable} shows the results from pairwise significance tests based on correlations before and after removing spatial effects using ESF. Without applying ESF the tests conclude significant association between each of the two sensitive features and bikeshare station numbers in a census tract. However, spatial autocorrelation completely accounts for this association, and the ESF-adjusted features do not show any significant correlation. 




\section{Conclusion}
\label{sec:conc}

We have shown that the influence of a common spatial factor may have a confounding effect on the estimation and evaluation of any association between a demographically sensitive feature and an outcome of interest. Our proposed hypothesis testing methods provide a first set of quantitative tools to tackle this problem. Spatial equity and accessibility has received attention in the geospatial data analysis literature \citep{ASHIK202077,Cobb}. But this attention has not yet carried over to the ML fairness community. We aim to bridge this gap through our paper, and foster new developments in this very relevant domain.



\bibliography{paper}

\appendix
\section{Preliminary concepts}

\subsection{Hypothesis testing}
\label{app:hyp}
In statistical hypothesis testing, the null hypothesis ($H_0$) represents a default assumption on the data generating process, which is tested against the alternative hypothesis ($H_a$). Given sample datas, a {\it test statistic} $T$ is calculated, and if the statistic value falls outside an interval, the test rejects $H_0$ in favor of $H_a$. Hypothesis tests can be seen as a decision function $\phi(\cdot)$ that takes a dataset $\cD$ as input, and given a significance level $1-\alpha, \alpha \in (0,1)$, along with hypotheses $H_0, H_a$, returns the decision whether to reject $H_0$ in favor of $H_a$ or not. Two types of errors can happen in this decision problem: (1) Type-I error, or the error of rejecting $H_0$ based on the data when it is true in reality, and (2) Type-II error, or the error of failing to reject $H_0$ when it is actually false. A good test would ideally have both errors small. Generally, in hypothesis problems a hard constraint on the upper bound of the type-I error is imposed:
$ \text{Pr} \left[ \phi(\cD; \alpha) = \text{Reject } H_0 | H_0 \text{ is true} \right] \leq \alpha, $
where the probability is over the randomness of the data.

A $p$-value refers to the probability that $T$ is at least as extreme (as high or low, depending on what $H_a$ is) as $t$---the value of $T$ for the sample data---given that $H_0$ is true. When $p$-value is $\leq \alpha$, we take the decision $\phi(\cD; \alpha) = \text{Reject } H_0$, else $\phi(\cD; \alpha) = \text{do not reject } H_0$. To calculate the $p$-value, we need access to the null distribution of $T$. Depending on the test, this is obtained either using known asymptotic results, or approximated using permutation or bootstrap.

\subsection{Bias metrics}
\label{app:metrics}
In algorithmic fairness, there are a number of bias detection metrics, such as disparate impact, demographic parity, and statistical parity \citep{MehrabiEtal19}. Many of these were proposed keeping classification problems in mind, work for binary sensitive attribute $A$, and either discrete labels $Y$ or probability outputs $\hat Y$. For example, the disparate impact (DI) metric in Section~\ref{sec:intro} calculates the ratio of positive probabilities given different values of $A$.

When one or more of $Y,A$ is continuous, statistical tests like Kolmogorov-Smirnov (KS) statistic, or $t$-tests can be used to distinguish between the distributions of the continuous attribute, conditioned on different values of the discrete attribute. When both $Y$ and $A$ are continuous, methods to correlate samples of continuous random variables, such as the Pearson's (linear) correlation coefficient and Spearman's rank correlation can be used.

\section{Implementation details}

\subsection{Permutation test}
\label{app:permtest}
The steps are as follows:
\begin{itemize}[leftmargin=*]
\setlength\itemsep{0em}
\item Suppose $\Pi_n$ is the set of all permutations of $\{1, \ldots, n \}$. For some permutation $\pi \in \Pi_n$, we calculate the Moran's I statistics from permuted samples, say $I_{y,\pi}, I_{j,\pi}$, and using these the P3 test statistics $T_{1,\pi}, T_{2,\pi}$.

\item  Given some large integer $M$, we obtain permuted samples of the test statistics under their respective null hypotheses by repeating the above for random permutations $\pi_1, \ldots, \pi_M$:
\begin{align*}
\mathcal T^0_1 = \{ T_{1,\pi 1}, \ldots, T_{1,\pi M} \};
\mathcal T^0_2 = \{ T_{2,\pi 1}, \ldots, T_{2,\pi M} \}.
\end{align*}

\item The permutation $p$-values are empirical tail probabilities of $T_1, T_2$ corresponding to the respective (empirical) null distributions:
\begin{align*}
\hat p_1 = \frac{1}{M} \sum_{m=1}^M \mathbb{I} (T_{1,\pi m} \geq T_1),
\hat p_2 = \frac{1}{M} \sum_{m=1}^M \mathbb{I} (T_{2,\pi m} \geq T_2).
\end{align*}

\end{itemize}
Note that to generate the joint null distribution of $(T_1,T_2)$, it is important to use the {\it same} set of permutations $\pi_i$ to generate null samples $(\mathcal T^0_1, \mathcal T^0_2)$ above \citep{ArborettiEtal18}.

\subsection{Details of ESF adjustment}
\label{app:esf}
When the feature being analyzed is continuous, we obtain its decorrelated version using residuals from its spatial autoregression
\begin{align*}
    (\hat \bfgamma_k, \hat \bfbeta) = \text{argmin}_{\bfgamma,\bfbeta}
    \left[ \frac{1}{2n} \| \by - \rho \Eb \bfgamma - \Xb \bfbeta \|^2 + \lambda \| \bfgamma \|_1^2 \right],\quad
    \tilde \by = \by - \rho \Eb_k \hat \bfgamma_k + \Xb \hat \bfbeta.
\end{align*}
When the feature being analyzed is discrete, we use an autologistic model \citep{autologistic}, which is simply a logistic version of the spatial autoregression model discussed. Thus, we estimate $\bfgamma, \bfbeta$ from the below model, again using $\ell_1$-penalization to select the number of eigenvectors of $\Eb$:
\begin{align*}
    p_{yi} := \text{Pr} (y_i = 1 | \Eb, \by, \Xb)
    = \left[ 1 + \exp \{ - (\rho \Eb \bfgamma + \Xb \bfbeta) \} \right]^{-1} ;\quad
    i = 1, \ldots, n.
\end{align*}

There are now three possible cases: both $Y,A$ are continuous, both are discrete, or one of them is continuous and the other is discrete. Below we describe ESF-adjusted testing strategies for each of these cases. In each case, we denote a generic bias metric computed before and after ESF by $r(\cdot,\cdot)$.

\paragraph{Case I: both continuous}
Examples of relevant bias metrics in this situation are Pearson's correlation coefficient, Spearman's rank correlation, and Kendall's Tau. In this case, the post-ESF bias metric is computed by simply using the respective residuals, i.e. $r(\tilde \by, \tilde \ba)$ in place of $r(\by, \ba)$, and compared against standard thresholds for that metric.

\paragraph{Case II: both discrete}
Disparate impact, equalized odds, and equality of opportunity are examples of bias metrics applicable when both $Y,A$ are discrete. In our setting, we first estimate the sample-level probabilities using ESF (
$i = 1, \ldots, n$):
\begin{align*}
    \hat p_{yi} = \left[ 1 + \exp \{ -(\rho \Eb_{ky} \hat \bfgamma_{ky} + \Xb \hat \bfbeta_y)\} \right]^{-1},\quad
    \hat p_{ai} = \left[ 1 + \exp \{ -(\rho \Eb_{ka} \hat \bfgamma_{ka} + \Xb \hat \bfbeta_a)\} \right]^{-1}.
\end{align*}
Following this, we generate $\hat y_i^b \sim \text{Bernoulli} (\hat p_{yi}), \hat a_i^b \sim \text{Bernoulli} (\hat p_{ai})$, and calculate $r(\hat \by^b, \hat \ba^b)$. We repeat this $B$ times, and obtain the collection of metrics $\{ r(\hat \by^b, \hat \ba^b): b =1, \ldots, B \}$---fixing $B=1000$ in our experiments. This simulates the empirical null distribution of the bias metric, i.e. under the assumption that the only bias present in the data is due to $\Wb$. We now simply compare the unadjusted metric against this distribution to obtain an approximate $p$-value: $ \hat p = B^{-1} \sum_{b=1}^B \BI( r(\hat \by^b, \hat \ba^b)
\geq r(\by,\ba))$.
%


\paragraph{Case III: one discrete, one continuous}
The KS statistic can be used as bias metric when one of $Y,A$ is discrete. In this situation, we simply estimate the sample class probabilities for the discrete feature (without loss of generality, $Y$) and approximate the empirical null distribution of bias metrics using multiple simulated copies of the ESF-adjusted $\tilde \by$ and the fixed, continuous fitted values $\hat \ba$: $\{ r(\hat \by^b, \hat \ba): b =1, \ldots, B \}$. The $p$-value is simply the tail probability at $r(\by,\ba)$ with respect to this empirical distribution.


\end{document}